\documentclass[manuscript]{aastex}

\slugcomment{To appear in ApJL}

\shorttitle{MHD explanation of counter rotation}
\shortauthors{Sauty et al.}

\begin{document}

\title{Counter-rotation in magneto-centrifugally driven jets and other winds
}

\author{C. Sauty and V. Cayatte}
\affil{Laboratoire Univers et Th\'eories, Observatoire de Paris, UMR 8102 du
  CNRS, Universit\'e Paris Diderot, F-92190 Meudon,  France}
\email{christophe.sauty@obspm.fr}

\author{J.J.G. Lima\altaffilmark{1}}
\affil{Centro de Astrof\'{\i}sica, Universidade do Porto, Rua das Estrelas,
  4150-762 Porto, Portugal}

\author{T. Matsakos\altaffilmark{2}}
\affil{CEA, IRAMIS, Service Photons, Atomes et Mol\`ecules, F-91191
  Gif-sur-Yvette, France}

\and

\author{K. Tsinganos}
\affil{IASA and Section of Astrophysics, Astronomy \& Mechanics, Department of
  Physics, University of Athens, Panepistimiopolis GR-157 84, Zografos, Greece}

\altaffiltext{1}{Departamento de F\'{\i}sica e Astronomia, Faculdade de
  Ci\^{e}ncias, Universidade Porto, Rua do Campo Alegre, 687, 4169-007 Porto,
  Portugal}

\altaffiltext{2}{IASA and Section of Astrophysics, Astronomy \& Mechanics,
  Department of Physics, University of Athens, Panepistimiopolis GR-157 84,
  Zografos, Greece}

\begin{abstract}
Rotation measurements in jets from T Tauri stars is a rather difficult task.
Some jets seem to be rotating in a direction opposite to that of the underlying
disk, although it is not yet clear if this affects the totality or part of the
outflows.
On the other hand, Ulysses data also suggest that the solar wind may rotate in
two opposite ways between the northern and southern hemispheres.
{We show that this result is not surprising as it may seem and that it emerges
naturally from the ideal MHD equations.
Specifically, counter rotating jets do not contradict the magneto-centrifugal
driving of the flow nor prevent extraction of angular momentum from the disk.}
{The demonstration of this result is shown by combining the ideal MHD
equations for steady axisymmetric flows.}
{Provided that the jet is decelerated below some given threshold beyond the
Alfv\'en surface, the flow will change its direction of rotation locally or
globally.
Counter-rotation is also possible for only some layers of the outflow at
specific altitudes along the jet axis.}
{We conclude  that the counter rotation of winds or jets with respect to the
source, star or disk, is not in contradiction with the magneto-centrifugal
driving paradigm.
This phenomenon may affect part of the outflow, either in one hemisphere, or
only in some of the outflow layers.
From a time dependent simulation, we illustrate this effect and show that it may
not be permanent.}
\end{abstract}

\keywords{ Magnetohydrodynamics (MHD); Stars: pre-main sequence; Stars: winds, outflows;
  Stars: mass-loss; Stars: rotation;  ISM: jets and outflows}
  
  \section{Introduction}

Many attemps have been made to measure jet rotation in various wavelengths
(e.g. \citeauthor{Coffeyetal04}, \citeyear{Coffeyetal04}), which is a rather delicate
and difficult task.
The RW Aur jet, among others, has been the most extensively studied such case.
\citeauthor{Cabritetal06} (\citeyear{Cabritetal06}) and
\citeauthor{Coffeyetal04} (\citeyear{Coffeyetal04}) conclude that the rotation
of the receding optical jet is opposite to that of the underlying disk.
However, observations in the near UV do not confirm this result for the
approaching jet \citep{Coffeyetal12}.
This clearly shows the difficulties of rotation measurements which require a
very precise experimental procedure.

Even though the RW Aur jet is not a convincing case, counter rotation has been
observed in several other jets.
For instance, there is at least one knot (SK1) in HH212 that seems to counter
rotate as explained in \cite{Coffeyetal11}, with new evidence coming from Fe and
HII lines.
This is also the case for HH111 whose observations indicate a counter rotating
knot too.

Moreover, observations of the solar wind by Ulysses also show that in situ
rotation measurements are delicate.
\cite{Sautyetal05} have plotted in Fig. 1 of their paper, the latitudinal
variation of the rotation velocity of the flow, the absolute value of which is
very small.
Although the conclusion is not definitive and could be due to some experimental
uncertainties, the plot may also indicate that the solar wind in the northern
hemisphere rotates in an opposite direction as compared to the wind in the
southern hemisphere.
Namely, the southern outflow may well be counter rotating with respect to the Sun.

Despite the fact that rotation measurements in jets is a difficult task, counter
rotation remains still an important open issue.
First of all because according to the common view this may contradict the
classical magneto-centrifugal outflow driving proposed in \cite{BlandfordPayne82} for
jet launching from a Keplerian disk.
The present letter aims at showing that this is not the case.
Counter rotation is a direct consequence of the velocity variation along the
flow as well as of the flux tube geometry.
It is shown that the conservation laws are satisfied and specifically that
angular momentum flux is constant along the flow.
This theoretical result is illustrated by a simulation which shows that non steady effects can also cause counter rotation.

\section{Governing equations for ideal steady axisymmetric MHD outflows}
\label{sec.2}

\subsection{Summary of the basic assumptions}

The basic equations governing ideal MHD outflows  are the  momentum, mass and
magnetic flux conservation, the frozen-in law for infinite conductivity and the
first law of thermodynamics.
In steady and axisymmetric conditions \citep{Tsinganos82, HeyvaertsNorman89}
there are five integrals, quantities that are conserved along a given (poloidal)
stream/field line.
These are the magnetic flux $A$, the mass to magnetic flux ratio, the energy { flux} $E(A)$,
the angular momentum { flux} $L(A)$  and the angular frequency or co-rotation frequency
$\Omega(A)$.
In the following, we denote cylindrical coordinates with ($z, \varpi, \varphi$)
and  spherical coordinates with ($r, \theta, \varphi$).

The total angular momentum flux and the co-rotation frequency can be expressed in
terms of the azimuthal and poloidal components of the magnetic and velocity
fields.
The toroidal component of the induction equation combined with the frozen flux
condition gives the co-rotation frequency $\Omega$,
\begin{equation}\label{Omega}
  \Omega(A)= \frac{1}{\varpi}
  \left(  V_\varphi- \frac{V_{\rm p}}{B_{\rm p}}B_\varphi  \right)
  \,.
\end{equation}
The angular momentum flux is obtained by integrating the momentum equation in the
azimuthal direction,
\begin{equation}\label{L}
  L(A)= \varpi
  \left( V_\varphi - \frac{B_{\rm p}}{4\pi\rho V_{\rm p}}B_\varphi  \right)
  \,.
\end{equation}
The poloidal Alfv\'en speed and the cylindrical Alfv\'en radius are defined as,
\begin{equation}
\label{varpi_star}
  V_{\rm A}^2 = \frac{B^2_ {\rm p}}{4 \pi \rho}\,, \varpi_{*}^2 = \frac{L}{\Omega}
\,.
\end{equation}
For a smooth crossing of the Alfv\'en surface, where $V_{\rm p}=V_{\rm A}$,
the cylindrical radius along a given fieldline must
adjust to this value, $\varpi = \varpi_{*}$.

From Eqs. (\ref{Omega})-(\ref{varpi_star}) we derive the azimuthal
components as functions of the free integrals and the poloidal components of the
fields, using the Alfv\'en speed and radius:
\begin{equation}\label{VphiBphi}
  V_{\varphi} =\frac{L}{\varpi}
  \frac{\frac{\displaystyle  V_{\rm p}^2}{\displaystyle V_{\rm A}^2}
  - \frac{\displaystyle \varpi^2}{\displaystyle \varpi_*^2}}
  {\frac{\displaystyle V_{\rm p}^2}{\displaystyle V_{\rm A}^2}- 1}
  \,,\qquad
  \frac{B_{\varphi}}{\sqrt{4\pi\rho}}
  =-\frac{L}{\varpi}\frac{V_{\rm p}}{V_{\rm A}}
  \frac{\frac{\displaystyle \varpi^2}{\displaystyle \varpi_*^2}-1}
  {\frac{\displaystyle  V_{\rm p}^2}{\displaystyle V_{\rm A}^2}-1}
  \,.
\end{equation}
With these expressions we show that a reversal of the rotation velocity is
possible.
Note that the toroidal components depend only on the poloidal ones and the free
MHD integrals, being independent of the thermodynamics of the flow.

\subsection{Reversal of the toroidal velocity}

\begin{figure}
\includegraphics[width= 16 cm] {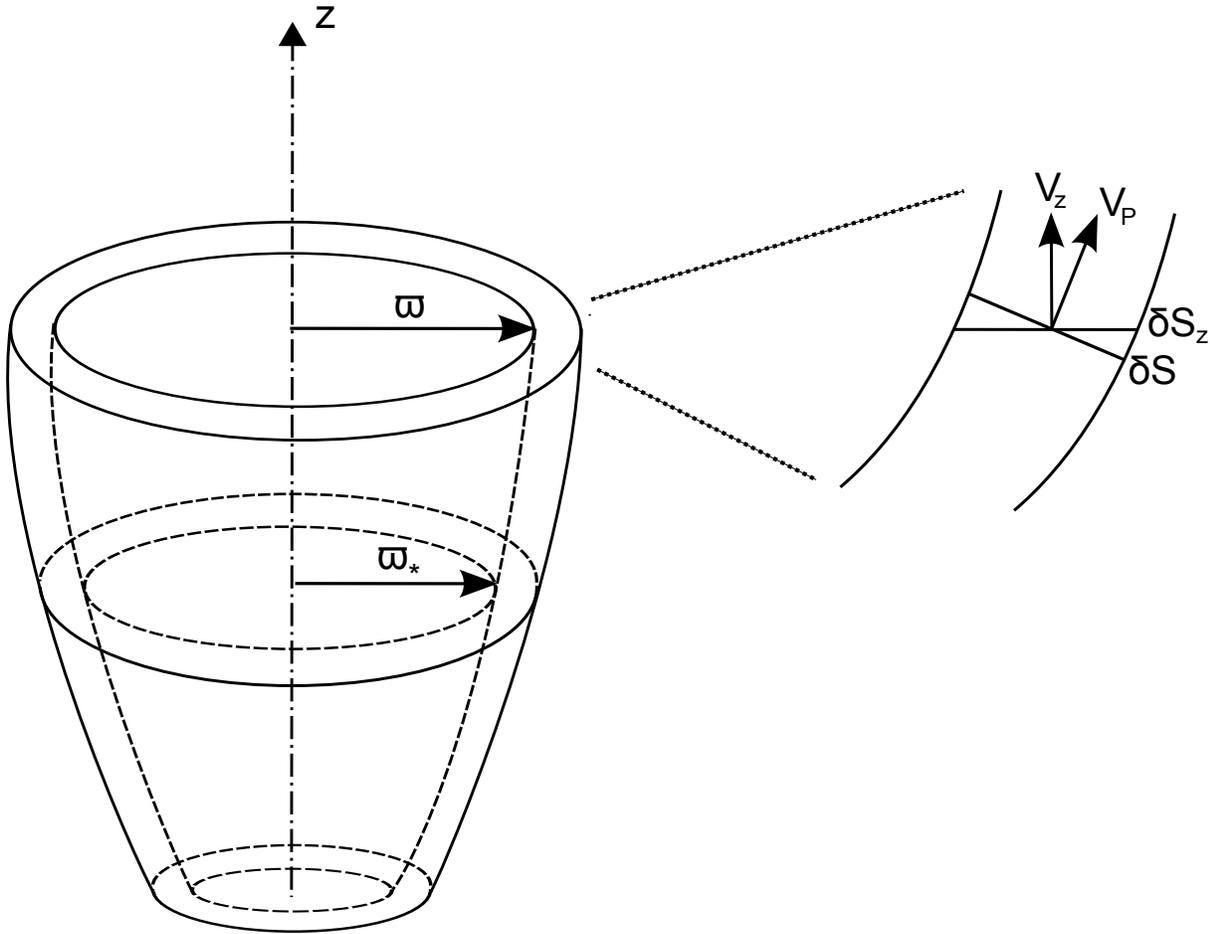}
\caption{3D view of a flux tube along the flow, where the cylindrical radius 
  is shown at the Alfv\'en transition and another point downstream. 
  On the right part of the figure, a zoom of the poloidal plane shows the
  definition of $V_{\rm p}$, $\delta S$, $V_z$ and $\delta S_z$.
}
\label{fig1}
\end{figure}

Consider the flow along a given stream/field-line as seen in Fig. \ref{fig1}.
Assuming that the flow remains everywhere super-Alfv\'enic after crossing the
corresponding critical surface, the denominator of Eq.~(\ref{VphiBphi}) is
always positive,
${\frac{\displaystyle  V_{\rm p}^2}{\displaystyle V_{\rm A}^2}- 1} > 0$.
The sign of the toroidal velocity is then given by the numerator,
\begin{equation}\label{Vphisign}
  {\rm sgn}(V_{\varphi}) = {\rm sgn}
  \left({\frac{\displaystyle V_{\rm p}^2}{\displaystyle V_{\rm A}^2}
  - \frac{\displaystyle \varpi^2}{\displaystyle \varpi_*^2}}\right)
  \,.
\end{equation}

The initially positive rotation velocity will change sign if the velocity of the
flow along the flux tube is less than the threshold value,
\begin{equation}\label{Vphithreshold1}
  V_{\rm p}< V_{\rm A}\frac{\displaystyle \varpi}{\displaystyle \varpi_\star}
  = V_{\rm threshold}
  \,.
\end{equation}
Note that this value is always above the local Alfv\'en speed because we expect
$\varpi>\varpi_\star$.

Along a given flux tube, we can write the magnetic and mass flux conservation as
\begin{equation}\label{Bflux}
  B_{\rm p}\delta S= B_{\star}\delta S_{\star} = {\rm constant}
\end{equation}
and
\begin{equation}\label{massflux}
  \rho V_{\rm p}\delta S= \rho_{\star} V_{\star}\delta S_{\star}= {\rm constant}
  \,.
\end{equation}
Note that $\delta S$ is the surface element perpendicular to the poloidal
velocity (see Fig. \ref{fig1}), such that one can always replace
$\vec{V}_{\rm p}$ by any of its components using the following equation,
\begin{equation}\label{massflux2}
  \vec{V}_{\rm p}\cdot \delta \vec{S} = V_{\rm p}\delta S= V_{z}\delta S_{z}
  = V_{r}\delta S_{r}= V_{i}\delta S_{i}
  \,,
\end{equation}
where $V_{i}$ is the $i$-th component of the velocity and $\delta S_{i}$ is the
cross section perpendicular to the $i$-th direction.
The calculation is simpler if one works with the poloidal velocity.
Nevertheless, it is more straightforward to compare the $z$ component of the
velocity and the cross section $\delta S_{z}$, which is in the $\varpi$
direction, since these are the observed quantities.

Using these two equalities in the numerator of the azimuthal velocity we
successively get:
\begin{eqnarray}\label{Vphidem1}
  \frac{\displaystyle  V_{\rm p}^2}{\displaystyle V_{\rm A}^2}
  - \frac{\displaystyle \varpi^2}{\displaystyle \varpi_{\star}^2}
  = \frac{\displaystyle 4\pi\rho V_{\rm p} V_{\rm p}}{\displaystyle B_{\rm p}^2}
  - \frac{\displaystyle \varpi^2}{\displaystyle \varpi_{\star}^2}
  = \frac
  {\displaystyle 4\pi \rho_\star V_{\star} \frac{\displaystyle \delta S_{\star}}
  {\displaystyle \delta S} V_{\rm p}}
  {\displaystyle B_{\star}^2 \frac{\displaystyle \delta S_{\star}^2}
  {\displaystyle\delta S^2 }}
  - \frac{\displaystyle \varpi^2}{\displaystyle\varpi_{\star}^2 }
  \,.
\end{eqnarray}
Since at the Alfv\'en surface we have $4\pi \rho_\star V_{\star}^2=B_{\star}^2$,
the above expression simplifies further,
\begin{equation}
  \label{Vphidem2}
  \frac{\displaystyle  V_{\rm p}^2}{\displaystyle V_{\rm A}^2}
  - \frac{\displaystyle \varpi^2}{\displaystyle \varpi_{\star}^2}
  = \frac{\frac{\displaystyle \delta S_{\star}}{\displaystyle\delta S}V_{\rm p}}
  {\frac{\displaystyle \delta S_{\star}^2}{\displaystyle\delta S^2 }V_{\star}}
  - \frac{\displaystyle \varpi^2}{\displaystyle \varpi_{\star}^2}
  = \frac{\displaystyle \delta S}{\displaystyle \delta S_{\star}}
  \frac{\displaystyle V_{\rm p}}{\displaystyle V_{\star}}
  - \frac{\displaystyle \varpi^2}{\displaystyle \varpi_{\star}^2}
  \,.
\end{equation}
Thus, the sign of this last expression determines the sign of the azimuthal
velocity in the super Alfv\'enic region:
\begin{equation}\label{Vphisign}
  {\rm sgn}(V_{\varphi}) =
  {\rm sgn}\left(\frac{\displaystyle V_{\rm p}}{\displaystyle V_{\star}}
  - \frac{\displaystyle \varpi^2}{\displaystyle \varpi_{\star}^2}
  \frac{\displaystyle \delta S_{\star}}{\displaystyle \delta S}\right)
  \,.
\end{equation}
This gives a new expression for the threshold value defined in
Eq.~(\ref{Vphithreshold1}),
\begin{equation}\label{Vphireverse}
  V_{\rm threshold}
  = V_{\star}\frac{\displaystyle \varpi^2}{\displaystyle \varpi_\star^2}
  \frac{\displaystyle \delta S_{\star}}{\displaystyle \delta S}
  \,.
\end{equation}

\begin{figure}
\includegraphics[width= 16 cm] {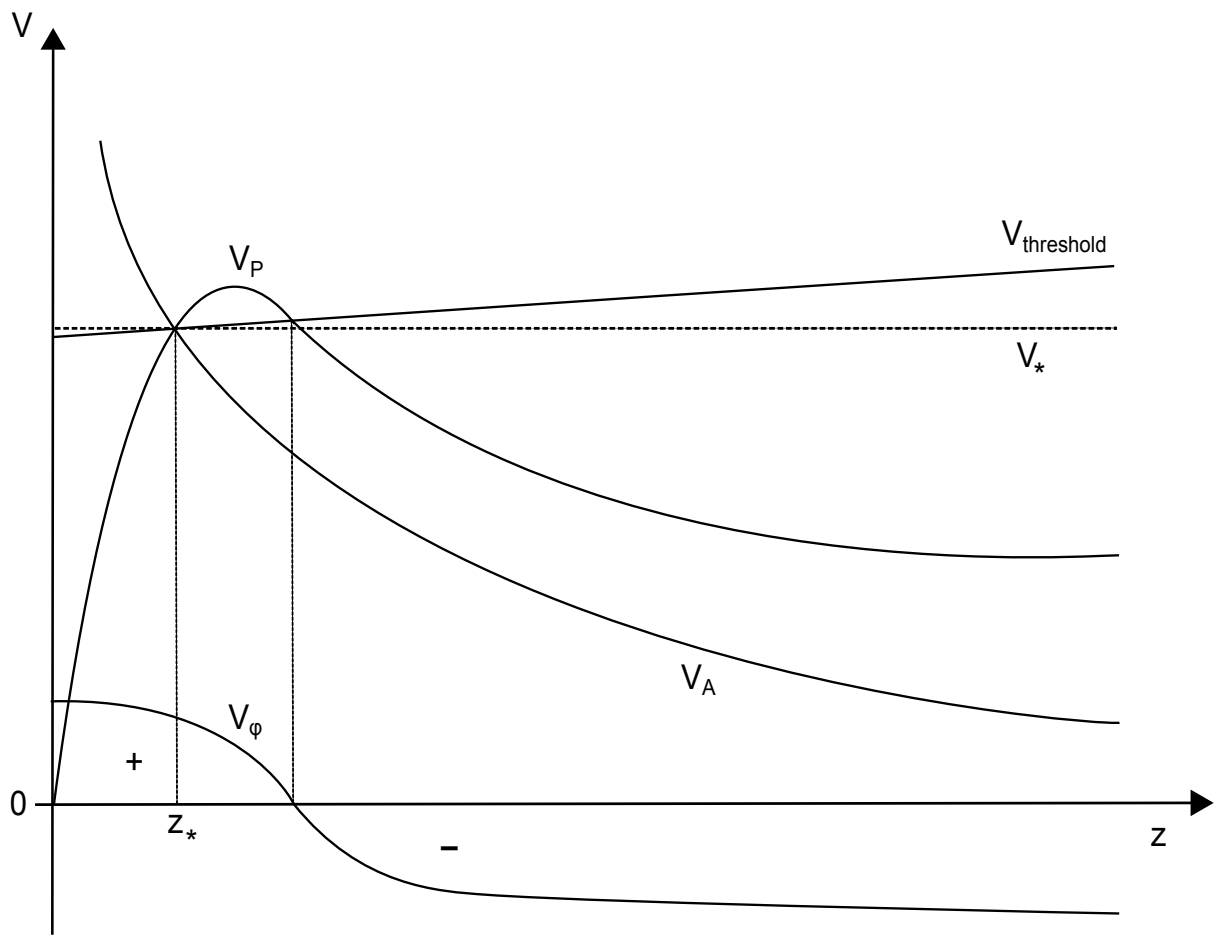}
\caption{Sketch along a given streamline showing the evolution of various
  velocities: the poloidal component $V_{\rm p}$, the threshold value
  $V_{\rm threshold}$, the poloidal Alfv\'en speed $V_{\rm A}$ and the rotation
  speed $V_\varphi$.
  The threshold value varies due to geometrical effects but remains close to
  $V_\star$ (dotted line), which is the value of the poloidal speed at the Aflv\'en point.
  The rotation changes direction at the location where the poloidal velocity far
  from the launching region becomes lower than the threshold value.
  Note that the flow remains super-Alfv\'enic after the Alfv\'en point.}
\label{fig2}
\end{figure}

If the flow remains super Alfv\'enic, reversal of the rotation takes place when
the velocity drops below some threshold value.
The flow can decelerate either because it expands and this provokes adiabatic
cooling, or because its kinetic energy drops due to some other mechanism such as
radiative losses.
The reversal might also occur if the threshold value increases.
This may be the case when the cross section of the tube decreases significantly
while the flux tube itself widens.
This is illustrated in Fig.~\ref{fig2}, namely, the toroidal velocity becomes
negative at the location where the poloidal speed curve crosses the line of the
threshold velocity.

Counter rotation is also obviously obtained if the flow becomes sub-Alfv\'enic.
In this case, the denominator of Eq. (\ref{VphiBphi}) becomes negative and the
reversal of the toroidal velocity is accompanied by a reversal of the toroidal
magnetic field too.
This may happen after a shock, which is probably similar to the suggestion of
\cite{Fendt11}, that the reversal of the toroidal field is due to the presence
of shocks.

Last, we also note that in the sub-Alfv\'enic part of the flow close to the source,
rotation may also change sign if there is a strong local acceleration and if the velocity increases above the threshold value (instead of below in the super-Alfv\'enic part).

Another way of exploring the counter rotation possibilities is to express the
numerator of the rotation speed from Eq.~(\ref{VphiBphi}) using the density
ratio,
\begin{equation}
  \label{Vphidem2}
  \frac{\displaystyle  V_{\rm p}^2}{\displaystyle V_{\rm A}^2}
  - \frac{\displaystyle \varpi^2}{\displaystyle \varpi_{\star}^2}
  =
  \frac{\displaystyle \rho_{\star}}{\displaystyle \rho}
  - \frac{\displaystyle \varpi^2}{\displaystyle \varpi_{\star}^2}
  \,.
\end{equation}
The sign of this quantity determines the sign of $V_\varphi$, i.e. reversal will
occur if $\rho>\rho_{\star}\frac{\varpi^2_{\star}}{\varpi^2}$.

\subsection{Discussion on the threshold value }

For meridionally self-similar models, it is more convenient to use the radial
component of the velocity, i.e. Eq.~(\ref{massflux2}). The criterion for a change of sign in the
 flow rotational velocity is now given by,
\begin{equation}\label{MSS}
V_{r}<V_{r,\star}\frac{\cos \theta}{\cos \theta_\star}
\,.
\end{equation}
This is particularly interesting for analytical stellar jet or wind models.

Similarly for radially self-similar models, reversal of the rotation occurs
where,
\begin{equation}\label{RSS}
  \vert V_{\theta} \vert < \vert V_{\theta,\star}
  \vert \frac{\sin \theta}{\sin \theta_\star}
  \,.
\end{equation}
This formula applies instead on analytical disk wind models.

If the physical quantities are uniform across the jet then $\delta S$ is
proportional to $\varpi^2 $.
The condition then reduces to the much simpler expression, $V_{\rm}<V_{\star}$.
The threshold value is exactly the Alfv\'en speed at the Alfv\'en surface.

To compare this criterion with observations and numerical simulations it is
useful to work with the vertical component of the velocity using
Eq.~(\ref{massflux2}) because this is the jet velocity component measured.
The surface element is then replaced by its projection along $z$, $\delta S_z$,
which is the cross section of the flux tube in the horizontal $\varpi$ direction
that corresponds to the jet radius as it is measured,
\begin{equation}\label{Vphireversez}
  V_{z} < V_{\star}\frac{\displaystyle \varpi^2}{\displaystyle \varpi_\star^2}
  \frac{\displaystyle \delta S_{z, \star}}{\displaystyle \delta S_z}
  \,.
\end{equation}

If we average the quantities at a given altitude $z$, we may have a criterion for the averaged velocity using the previous equation, namely,
\begin{equation}\label{Vphireversezaverage}
  {\bar V_{z}}
  = \frac{\displaystyle \int{V_{z} \delta S_z}}{\displaystyle \varpi^2} 
  < \frac{\displaystyle \int{V_{\star} \delta S_\star}}{\displaystyle \varpi_\star^2} 
  = {\bar V_{\star}}
  \,.
\end{equation}
The criterion (i.e. $\bar V_{\rm}<\bar V_{\star}$) is equivalent to the criterion for transversaly uniform flow.

This may be more adequate to analyse observational data and jet models such as the one proposed by \cite{Lovelaceetal91}. In fact,
Eq. (16) of \cite{Lovelaceetal91} shows that an increase of the average jet velocity induces a decrease of the rotational one, which could become negative. 

\subsection{Counter rotation in simulations}
\label{sec.3}

We have performed a numerical simulation in
which several reversals of the toroidal velocity are observed.
We have adopted the initial conditions of \cite{Matsakosetal12}, model NIPH,
in which two analytical solutions for the stellar and disk wind components are
combined and a constant external pressure is applied beyond a specific radius.
We have carried out the temporal integration of the MHD equations using PLUTO
\citep{Mignoneetal07}, in a resolution of $512\times2048$ up to a final time
that corresponds to $\sim$$100$ crossing times of the simulation box.

\begin{figure}[h!]
\includegraphics[width= 16 cm] {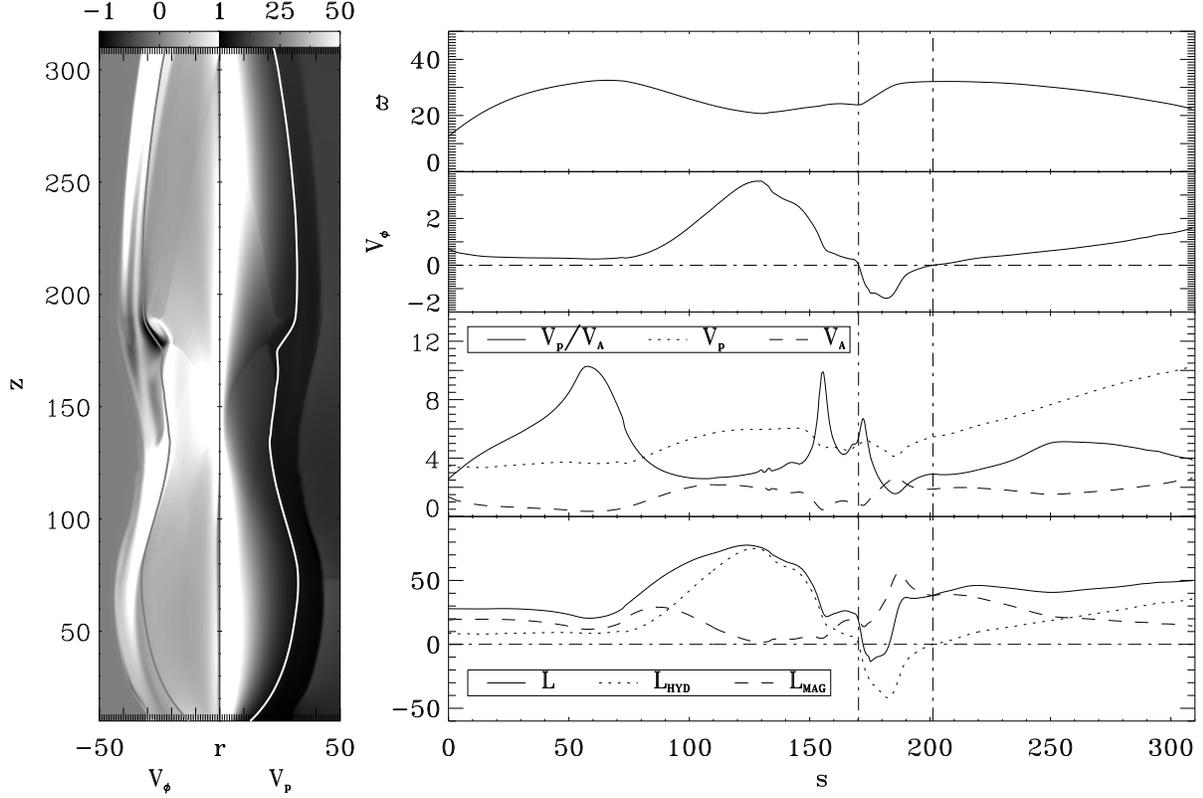}
\caption{Snapshot of a two component jet simulation surrounded by an external
  medium of constant pressure.
  On the left panel, the toroidal velocity (left) and the corresponding poloidal
  velocity (right) are displayed. { It also gives the cylindrical radius $\varpi(z)$ as a function of $z$.}
  The reversal of the rotation corresponds to the darkest zone of the
  $V_\varphi$ contours.
  A specific streamline has been selected (solid line) along which the
  quantities shown on the right panel are plotted.
  From top to bottom:  { the cylindrical radius $\varpi(s)$ the angular momentum flux, the toroidal speed, the poloidal and Alfv\'en 
  speeds, and   the poloidal Alfv\'enic Mach number as functions of the curvilinear abscissa, $s$. The angular momentum flux (solid 
  line) has two components, the hydrodynamical one 
  ($L_{\rm HYDRO}$, dotted line) and the magnetic one  ($L_{\rm MAG}$, dashed line). }
}
\label{fig3}
\end{figure}

During the initial transient phase, a strong shock propagates upwards due to the
fact that the initial state is not in equilibrium.
Together with the shock, a significant reversal of the toroidal velocity is also
observed, although this phenomenon is not directly related with the above
analysis.
Once the system reaches a quasi-steady state, several layers demonstrate an
opposite $V_\varphi$.
These counter rotating regions remain roughly at the same location for large
time intervals, before they vanish and reappear.
They obviously remain transient phenomena.

The left panel of Fig. \ref{fig3} displays contours of the toroidal velocity on
the left and the poloidal component on the right.
We have selected a streamline ($\varpi(z)$, grey and white solid lines of left panel) that crosses the region where the rotation
reverses, along which the cylindrical radius, the angular momentum fluxes, the toroidal, poloidal and poloidal Alfv\'en speeds as well 
as the poloidal Alfv\'enic Mach number are shown on the right panel of Fig. \ref{fig3}.

The plots suggest that the $V_\varphi$ component becomes negative at the
location where the poloidal velocity decreases approaching the Alfv\'en speed.
Note that the flow remains always super-Alfv\'enic.
The dotted dashed vertical lines indicate the change of sign of the rotation.
In this region, the poloidal velocity reaches the local threshold value which
varies along the flow.
Although the jet is not an exact steady flow, the simulation results can be
explained based on the analysis that describes the super Alfv\'enic regime.

On Fig. \ref{fig3} we plot the total angular momentum flux $L$ and its  two components (see Eq. \ref{L}), namely th
hydrodynamical part  $L_{\rm HYDRO}=\varpi V_\varphi $ and the magnetic one $L_{\rm MAG}
=-\varpi \frac{B_{\rm p}}{4\pi\rho V_{\rm p}}B_\varphi $. The total angular momentum flux is clearly not conserved because the 
configuration is not in an exact steady state. However the 
reversal of the toroidal velocity corresponds to an exchange between the two components where the magnetic part becomes 
dominant.  $L$ drops rapidly up to $s=190$. In this first part the reversal of the velocity is strongly time dependent and related to the 
shock as suggested by \cite{Fendt11}. Above $s=190$, $L$ is almost
constant and in this region our steady state criterion for the reversal is valid. It is a nice illustration of the above theory.

\section{Conclusions}
We have presented a simple study showing that counter rotation may occur
in any jet or wind independently of the launching mechanism.
Counter rotation may be observed in the outer layers of a disk wind, as shown in
our numerical simulation and is observed in some protostellar jets. 
It may also occur in a stellar wind, as suggested from the data of the solar
wind.
This effect does not contradict the disk launching mechanism, nor the
conservation of the angular momentum along the poloidal field lines.
The criterion that marks the reversal of $V_\varphi$ is determined by either a deceleration of
the flow or a specific geometrical configuration.

For the Sun, the toroidal velocity of the solar wind is very small or negative because there is
almost no acceleration after the Alfv\'en point.  
In outflows associated with YSOs  we expect that there may be reversals of the
rotation in some of the layers of the flow but not necessarily across or along
the whole jet.
There is no evidence that this phenomenon should be permanent even if it is not
a transient modification of the flow dynamics.
In all cases, the total angular momentum is conserved, Eq.~(\ref{L}), although
the kinetic and magnetic terms constituting $L(A)$ are expected to exchange
the amount they store.
The situation is similar to the transfer of energy from the enthalpy to the
kinetic term in the Bernoulli integral in the thermally driven Parker wind, or,
to the transfer of energy from the Poynting energy term at the base to the
kinetic energy term at large distances in the generalized Bernoulli integral in
magneto-centrifugal winds.

\acknowledgments

The authors are grateful to C. Fendt, F. Bacciotti, A. K\"onigl,
C. Stehl\'e, J.P. Chi\`eze for fruitful discussions. We also thank an anonymous referee for her/his valuable comments which help us to improve the manuscript.
This work has been partly supported by a Marie Curie European Reintegration
Grant within the 7th European Community Framework Programme (``TJ-CompTON; PERG05-GA-2009-249164'')
and partly by the project \lq\lq Jets in young stellar objects: what do simulations tell us?\rq\rq\ 
funded under a 2012 \'{E}gide/France-FCT/Portugal bilateral cooperation Pessoa Programme.

\clearpage

\end{document}